\documentclass[10pt,letterpaper]{article}
\usepackage[top=0.85in,left=2.75in,footskip=0.75in]{geometry}

\usepackage{changepage}

\usepackage[utf8x]{inputenc}

\usepackage{textcomp,marvosym}

\usepackage{amsmath,amssymb}

\usepackage{cite}

\usepackage{nameref,hyperref}

\usepackage[right]{lineno}

\usepackage{microtype}
\DisableLigatures[f]{encoding = *, family = * }

\raggedright
\usepackage[aboveskip=1pt,labelfont=bf,labelsep=period,justification=raggedright,singlelinecheck=off]{caption}

\makeatletter
\renewcommand{\@biblabel}[1]{\quad#1.}
\makeatother

\date{}

\usepackage{lastpage,fancyhdr,graphicx}
\usepackage{epstopdf}
\fancyheadoffset[L]{2.25in}
\fancyfootoffset[L]{2.25in}
\begin{document}
\vspace*{0.2in}

\begin{flushleft}
{\Large
\textbf\newline{Urban Attractors: Discovering Patterns in Regions of Attraction in Cities} 
}
\newline
\\
May Alhazzani\textsuperscript{1},
Fahad Alhasoun\textsuperscript{2},
Zeyad Alawwad\textsuperscript{1},
Marta C. Gonz\'{a}lez\textsuperscript{2,3},

\bigskip
\textbf{1} Center for Complex Engineering Systems, King Abdulaziz City for Science and Technology, Riyadh, Saudi Arabia
\\
\textbf{2} Department of Civil and Environmental Engineering, Massachusetts Institute of Technology, Cambridge, MA 02139.
\\
\textbf{3} Center for Advanced Urbanism, Massachusetts Institute of Technology, Cambridge, MA 02139.
\\
\bigskip

%
%


* martag@mit.edu

\end{flushleft}
\section*{Abstract}
Understanding the dynamics by which urban areas attract visitors is significant for urban development in cities. In addition, identifying services that relate to highly attractive districts is useful to make policies regarding the placement of such places. Thus, we present a framework for classifying districts in cities by their attractiveness to visitors, and relating Points of Interests (POIs) types to districts' attraction patterns. We used Origin-Destination matrices (ODs) mined from cell phone data that capture the flow of trips between each pair of places in Riyadh, Saudi Arabia. We define the attraction profile for a place based on three main statistical features: The amount of visitors a place received, the distribution of distance traveled by visitors on the road network, and the spatial spread of where visitors come from. We use a hierarchical clustering algorithm to classify all places in the city by their features of attraction. We detect three types of Urban Attractors in Riyadh during the morning period: {\emph Global} which are significant places in the city, {\emph Downtown} which the central business district and Residential attractors. In addition, we uncover what makes these places different in terms of attraction patterns. We used a statistical significance testing approach to rigorously quantify the relationship between Points of Interests (POIs) types (services) and the 3 patterns of Urban Attractors we detected. The proposed framework can be used for detecting the attraction patterns given by type of services related to each pattern. This is a critical piece of information to inform trip distribution models.


\section*{Introduction}
Understanding how different places in the city influence human mobility is significant for urban and transportation planning. A pressing need, in complex and congested cities is maintaining a robust transportation infrastructure. Understanding the patterns by which places in the city attract visitors is essential for planning and modifying the transportation system. Specifically, new data sources can help us to better predict the relation between population and trips attraction. This issue is of a particular significance to the city of Riyadh, Saudi Arabia where the largest metro project is being developed and promised to start running in 2019 \cite{ADA,louail2015uncovering}. Moreover, understanding how different types of places affects the flow of trips in the city differently helps to inform decisions and policies related to placing and modifying concentration of services. For example, how an industrial area influences the flow of trips during different times of the day; or where to place new business stores for higher profitability \cite{karamshuk2013geo}.

Today, with the ubiquity and pervasiveness of technology, data generated from mobile phones enable data analysts to better understand the behavior of individuals across many dimensions including their mobility patterns \cite{zheng2014urban,blondel2015survey}. An interesting area is how patterns of human mobility are affected by different places in the city.
The standard approach on categorizing urban areas classifies regions by their functionality and land use (i.e. commercial, educational, ...etc.). Recent works consider the human mobility aspects to classify regions. For example, Yuan {\it et al.} \cite{yuan2015discovering} proposed a topic modeling approach to classify districts into functional zones according to people's socioeconomic activities mined from taxi and public transport traces and points of interests (POIs) data. Pan {\it et al.} \cite{pan2013land} proposed a land use classification approach based on the social functions of districts also analyzed from GPS taxi traces where districts witness change of land use class dynamically. Toole {\it et al.} \cite{toole2012inferring} analyzed cell phone data to test cell phone activity patterns to classify for land use types. Less is known about how phone data can help to classify urban regions based on how attractive they are to different origins. 

In this work, we present a novel computational framework for classifying urban places by their attraction patterns. We define attraction profiles in terms of statistical features of incoming trips on a given time window. Different places in the city attract visitors differently. Some places like universities and hospitals attract a large amount of visitors who come from all over the city and travel long distances to go there. On the other hand, some places in the city that provide local services such as restaurants, schools, and small clinics only attract few people from nearby areas. We aim to automatically identify patterns of attraction of places based on three main dimensions: how many visitors a place receives, where visitors are coming from, and how long visitors are traveling to reach that place. Secondly, we further show what makes a region attraction behavior. To accomplish that, we used statistical significance testing to automatically relate the decomposition of POI types (services) and the discovered attraction patterns. This information is useful to relate type of businesses and attraction profiles.

The main contribution to this work is as follows:
\begin{itemize}
    	\item We present a computational framework for detecting attraction patterns and further relating POI types to each pattern of attraction.
	\item  We classify attraction patterns via the spatial dispersion of trip origins, the distribution of distances traveled by visitors through the road network and the total number of trips.
	\item We quantify the significance of POI types in a region using a statistical significance testing approach, which performs well in the context of phone and POI data.
\end{itemize}

\section*{Related work}
Multiple studies used human mobility behavior to classify urban areas. A recent study investigated the relationship between land use and mobility\cite{lee2015relating}. The authors showed that purposes of people's trips are strongly correlated with the land use of the trip's origin and destination. Recently, the availability of dynamic sources of data allowed for dynamic segmentation of the city according to human mobility behavior. Some studies combined human mobility with land use or POIs data to segment districts in urban areas according to their functions or use. The type of data used to capture human mobility behavior varies between individual GPS traces \cite{zheng2009mining,fan2014cityspectrum}, taxi pick up/drop off locations as in \cite{liu2015revealing, pan2013land} , Call Detail Records (CDRs) as in \cite{louail2015uncovering,toole2012inferring}, social media check ins as in \cite{zhan2014inferring,long2012exploring,bassolas2016touristic}, and bus smart card data as in \cite{han2015discovering}. This work differs from previous studies, by being the first to classify the urban regions through attraction profiles. 

Survey travel data has been used to detect the centers (significant places) of a city \cite{zhong2015revealing,de2016death}. A recent study proposed a method for measuring the centrality of locations that incorporates the number of people attracted to the location and the diversity of activities in which visitors engage \cite{zhong2015revealing}. The proposed method was tested on survey travel data in Singapore to identify the functional centers and track their significance over time. A similar approach focused on analyzing the aggregate behavior of the population to predicted highly attractive events such as the times square during new years count down in New York \cite{fan2015citymomentum}. Our method is based on validated Origin Destination (ODs) matrices mined from massive cell phone data that captures human mobility. More significantly, our approach incorporate not just the amount of people a place attracts, but also on where do they come from and the road distance they traveled.

Network analysis methods were used to detect hotspots based on flow patterns between locations\cite{louail2015uncovering,wu2016incorporating}. A recent paper \cite{louail2015uncovering} used ODs matrices extracted from cell phone data to identify the signature of mobility behavior as 4 main types of movements within the city: between hotspots, to hotspots, originating at hotspots and random flows. They showed how different cities have different mobility signatures. Additionally, a recent study used Taxi drop off/pick up traces in Shanghai to create a network of flow between places. They applied community detection to extract sub regions and analyze the interaction between and within sub regions. They found that urban sub-regions have larger internal interactions, while suburban centers are more significant on local traffic.%
This work made the breakdown of flow patterns instead of the impact of the place in attracting visitors, which is our aim in this paper.
%
Researchers adapted modeling approaches from Natural Language Processing (NLP) in identifying functional zones in urban areas \cite{yuan2015discovering,yuan2012discovering}. One study applied a Latent Dirichlet Allocation (LDA) model on Foursquare check-ins to detect local geographic topics that indicate the potential and intrinsic relations among the locations in accordance with users' trajectories. Additionally, a recent study used LDA and POIs to detect functional zones\cite{yuan2015discovering}. Our work is different where we aim to analyze the attraction behavior of a place using measures that has not been used in any of the previous work.

\section*{Urban Attractors Framework}
Fig.\ref{fig:0} shows the general structure of the process of analyzing attraction patterns in cities with the input datasets and the outputs. The first step in the process is to extract trips information from Call Detail Records (CDRs) of cell phones using the validated origin destination extraction algorithm implemented in \cite{toole2015path}. We use the ODs as a data source for estimating human mobility, where it provides the amount of trip from each pair of origin and destination. From the ODs, we mine three statistical features that quantify how attractive a place is: the number of trips a place receives, the spatial dispersion of the origins of all incoming trips, and the distance distribution visitors traveled to visit the place on the road network. We use these attraction features to classify all regions in the city according to their attraction behavior. Finally, using a statistical significance testing approach we relate each type of POIs that are significantly concentrated in each types of attractors identified. In the following sections we explain the process and its output in detail.

\begin{figure}[!h]
	\centering
	\frame{\includegraphics[width=0.7\columnwidth]{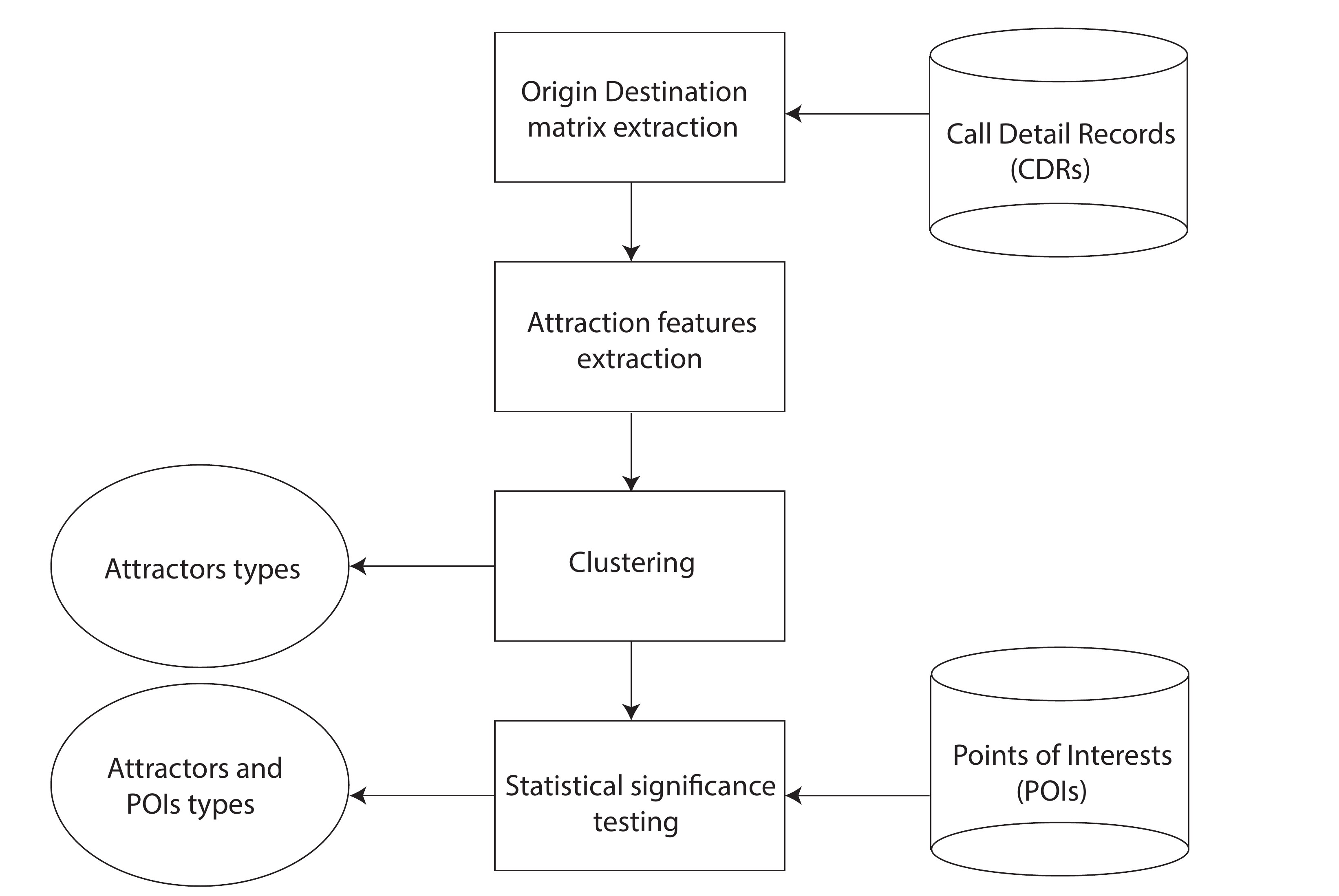}}
	\caption{Urban Attractors framework}~\label{fig:figure0}
\end{figure}


\section*{Origin Destination Matrix Extraction}

\begin{figure}[!h]
	\centering
	\frame{\includegraphics[width=0.5\columnwidth]{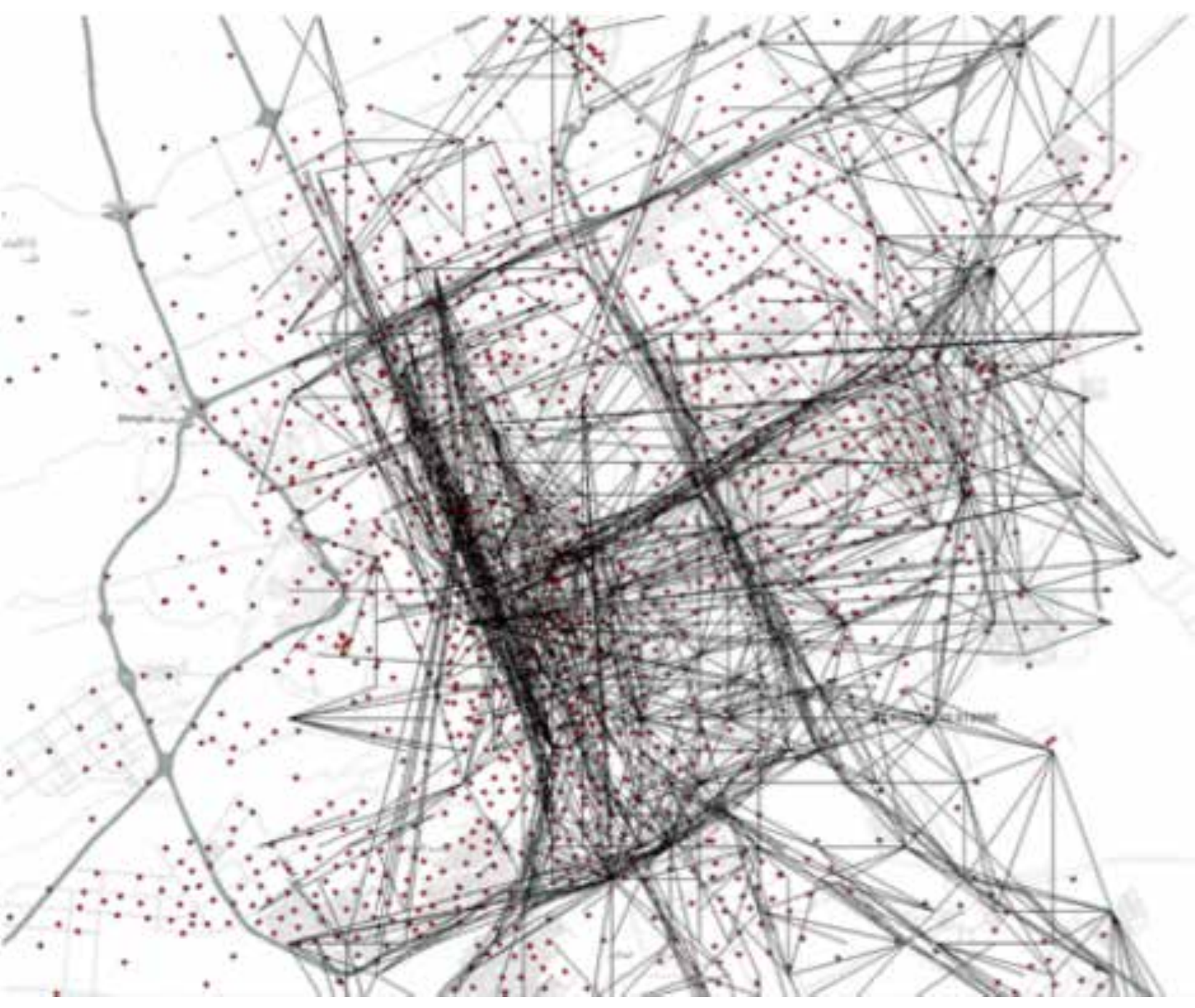}}
	\caption{The ODs in Riaydh during the morning period. Each line represents a trip from a source to a destination.}~\label{fig:figure1}
\end{figure}

The aim of this process is to extract the Origin-Destination matrices (ODs), that provides the number of trips between each pair of locations in the city for a specified time window of a typical weekday. Methods of estimating ODs range from traditional methods to more modern ones. Traditional methods include running surveys within cities and estimating the flows between locations of the city from the feedback of those surveys. Such methods consume longer periods of time and are inaccurate at times.  They usually span smaller population sample sizes and thus are more prone to biases. Recent research in the domain of ubiquitous computing provided alternative methodologies for estimating more accurate ODs from user generated datasets like cell phone data. The methods proposed in \cite{toole2015path,toole2014path} uses mobile phone location traces (i.e. CDRs) to estimate the flows of people between areas in the city. The large scale of the cell phone data provide sufficient sample sizes and more accurate information compared to traditional methods.  In this paper, we use state of the art methods of extracting OD matrices for the city of Riyadh between each pair of traffic analysis zones (TAZes) as shown in Fig.\ref{fig:figure1}.

 Our primary source of data is one month (December 2012) of Call Detail Records (CDRs) of anonymous mobile phone users in the city of Riyadh, Saudi Arabia. Within the CDRs, each record contains an anonymized user ID of the caller and receiver, the type of communication (i.e., SMS, MMS, call, data etc), the cell tower ID facilitating the service, the duration, and a time stamp of the phone activity. Each cell tower ID is spatially mapped to its latitude and longitude where each Voronoi cell in Fig.\ref{fig:figure1} correspond to a tower. The CDRs data contains more than 3 million unique users, which is a representative sample of Riyadh's population. Thus, the CDRs provide a proxy for tracking human mobility behavior in the city. However, computational steps are needed to extract clean trajectories from the CDRs.
 
The computational steps we take to transform raw mobile phone data (CDRs) to ODs are summarized as follows. First, we estimate transition probabilities between labeled locations such as home, work or other. Next, we filter users by number of records per day such that these locations are labeled with enough confidence (details of the method in ~\cite{toole2015path}).
A trip is marked by probable departures and arrivals to such locations within a specified time window. Then we scale from users to total population using census data. The output of this process is the OD matrix that indicates the number of individuals traveling between every possible pair of locations between 7 am and 10 am on a typical weekday.

 The spatial scale we used for locations is based on Traffic Analysis Zones, which is the official segmentation used in transportation planning. Conventionally segmenting the city into TAZes are based on census block information such as population per hour, where zones tend to be smaller in denser areas and larger in areas of low density. The TAZ based segmentation is more flexible and useful in analyzing places attraction patterns than using other segmentations such as neighborhood based or spatially uniform segmentation. Thus, we define our OD matrix $T$ by aggregating cell phone towers  on $1492$ TAZes. The elements of the matrix are the number of trips between each pair of TAZes $(i,j)$ in the city.


%

\begin{figure}
	\centering
	\includegraphics[width=0.7\columnwidth]{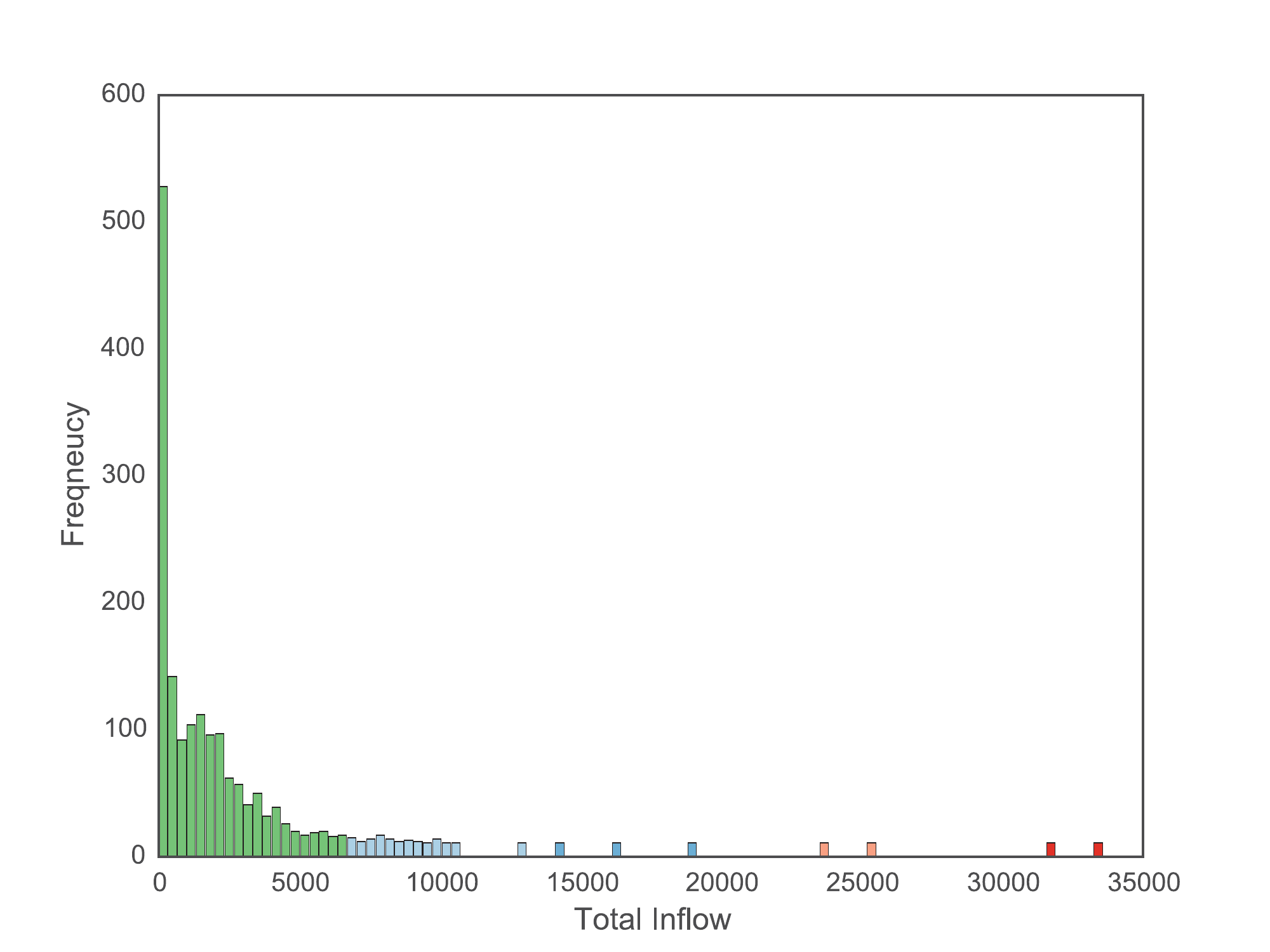}
	\caption{The distribution of total inflow received by TAZ's in Riaydh. the majority of places receive small to medium number of visitors. Few places receive very high inflow (colored red),which makes them highly attractive.}~\label{fig:figure2}
\end{figure}

\begin{figure*}[!t]
	\centering
	\frame{\includegraphics[width=1\columnwidth]{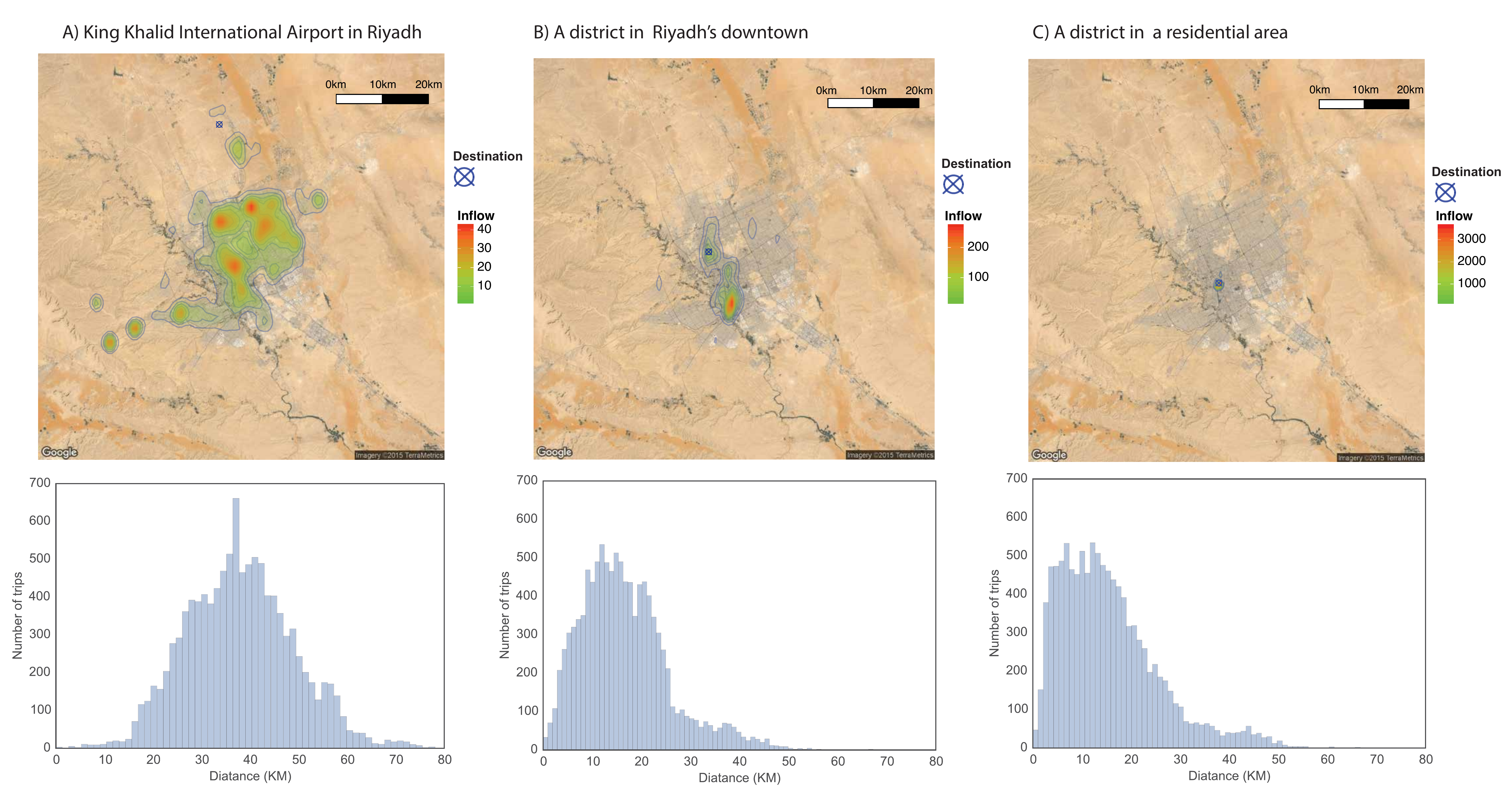}}
	\caption{Spatial dispersion and distance distributions of three examples of types of attractors (marked by black dots). (A) The international airport in Riaydh. (B) A place in the downtown area. (C) A place in a residential area. The top row shows heatmaps of the origins of the inflow, where the color differs corresponds by the amount of trips from that location. The bottom row is the distribution of road distance traveled by visitors of the selected place.}~\label{fig:figure3}
	
\end{figure*}

\section*{Attraction Features}\label{features}
We aim to identify different patterns of attraction through statistical features of the inflow to each place in the city. The first feature is the total amount inflow a place receives. As the more visitors a place receives, the more attractive that place is. Additionally, a place is more attractive if it attracts people from various places in the city. Some places only attract people nearby which makes them local in terms of from where they attract people. On the other hand, some place attracts people from all over the city such as universities  and hospitals. Thus, the second feature we measure is how spatially dispersed the origins are. Finally, we quantify the distances traveled to visit the place on the road network. In the following sections, we detail the calculations of each feature.

\subsubsection*{Inflow}
The amount of visitors a place receives is the strongest indicator of how attractive the place is. This feature measures the attraction force of a location , where locations that have high inflow (number of visitors) are major attractors in the city. Fig.\ref{fig:figure2} shows the distribution of the number of TAZes according to their inflow amount. The majority of TAZes have small to moderate inflow. However, there are few TAZes that have a very large inflow (colored red), which makes them highly significant. The inflow magnitude of a TAZ $i$ is simply calculated from the OD matrix as follows:
\begin{equation}
Inflow_i=\sum_{j=1}^{n} T_{ji}
\end{equation}

,Where $n$ is the total number TAZes, and $T_{ji}$ is the number of trips from TAZ $j$ to TAZ $i$.

\subsubsection*{Spatial dispersion}
Another way to measure the popularity of a place is to measure the spatial dispersion of the origins it attracts. The spatial dispersion quantifies how spatially dispersed the locations of the origins of trips are in relation to the center of mass of all origins. A place is more attractive if it attracts visors from various and spread-out places in the city. Major attractors tend to attract people from all over the city (large spatial dispersion), while less significant attractors only attract people nearby (presenting small spatial dispersion).

We measure the spatial dispersion of visitors by calculating a weighted standard distance deviation, which is a standard method used to measure the statistical dispersion of spatial data \cite{mitchell2005esri}. Mathematically, the weighted spatial dispersion (SD) for a TAZ $i$ is defined as follows:
\begin{figure*}[t]
	\centering
	\frame{\includegraphics[width=1\columnwidth]{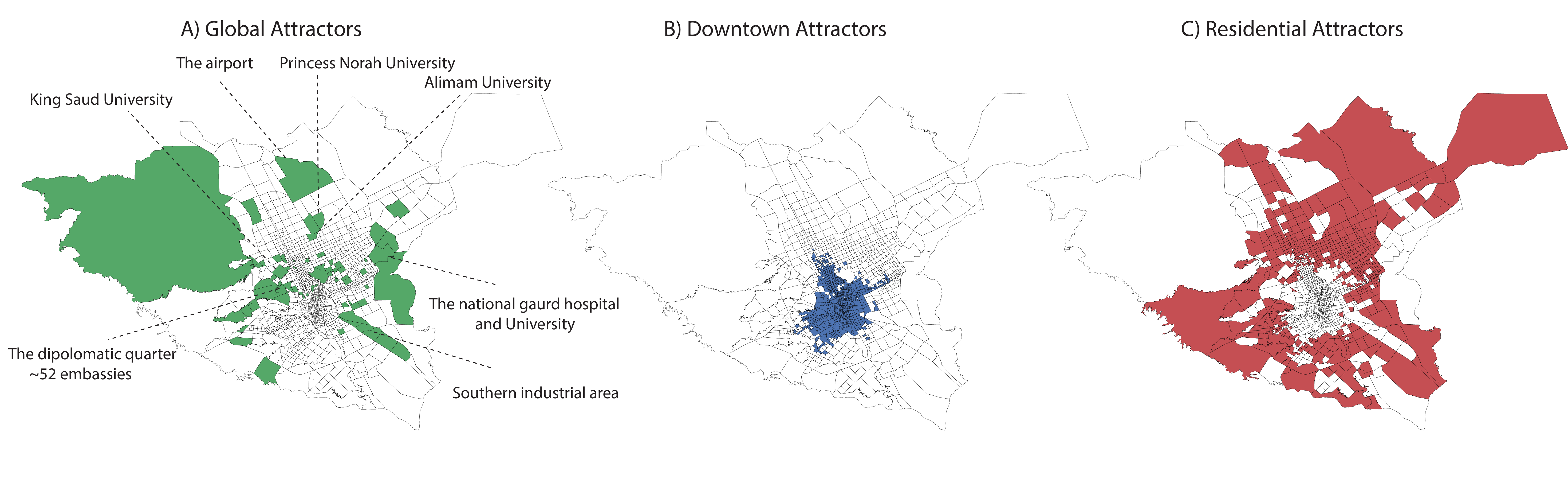}}
	\caption{Types of attractors. (A) Global at tractors, which are places with large number of visitors from all over the city traveling long distances to visit them. These include unique places in the city like the airport, large universities, hospitals, industrial areas, and the diplomatic quarter as annotated on the figure. (B) Attractors with large number of visitors due to their central location (downtown) they are more accessible and thus the distance visitors travel to reach them are shorter. (C) Less attractive places, that are located on the outer (residential) areas of the city.}~\label{fig:figure4}
	
\end{figure*}

\begin{equation}SD_i=\dfrac{ \sqrt{ \sum_{i=1}^{n}w_i(X_i-X_c)^2 + \sum_{i=1}^{n}w_i(Y_i-Y_c)^2} }{\sum_{i=1}^{n}w_i}
\end{equation}

Where $n$ is the total number TAZes. $X_i$ and $Y_i$ are the spatial coordinates of the origin of a trip $i$. $w_i$ is the amount of inflow from source TAZ $i$. $X_c$,$Y_c$ are the coordinates of the spatial center of mass of all origins of all the incoming flow calculated as follows:
\begin{equation}
X_c=\dfrac {\sum_{i=1}^{n}w_i.X_i}{\sum_{i=1}^{n}w_i} , Y_c=\dfrac {\sum_{i=1}^{n}w_i.Y_i}{\sum_{i=1}^{n}w_i}
\end{equation}

Fig.{\ref{fig:figure3}} shows three examples of places with different attraction behavior. The top row shows the heat maps of the inflow sources and their concentration. The destination TAZ is labeled with a target sign on the maps. Example $A$ shows the heat map of the international airport in Riyadh city, where the heat is spread all over the city, which indicates strong attraction. Example $B$ is a TAZ in the downtown area, where the inflow sources are moderately spread. Example $C$ is in a TAZ in a residential area, where it only attract visitors nearby with small spatial dispersion.

\subsubsection*{Distance distribution}
Another characteristic that defines the attraction patterns is the distribution of distances traveled by all the trips the destination receives. The trip distance from each source to the centroid of the attractor were calculated on the road network of Riyadh by using the Dijkstra shortest path algorithm \cite{dijkstra1959note} to find the optimal routes between all of the origin-destination pairs. This provides a more accurate estimation than the Euclidean or the Manhattan distances, as it accounts for the variation in the geometry of the road network. 

The bottom row in Fig.{\ref{fig:figure3}} shows the distance distribution of all trips a TAZ received. In Fig.{\ref{fig:figure3}} $A$ the distance distribution to the airport is unique, with long mean distance (around 40 km.) and a shift in the distribution due to the distant location of the airport in the very far north of the city. On the other hand, Fig.{\ref{fig:figure3}} $B$ has intermediate length of the mean distance with a distribution tail that corresponds to the long distance traveled by some visitors to reach downtown. Finally, in Fig.{\ref{fig:figure3}} $C$, most of the trips correspond to short distances. Clearly, for different types of attractors the distance distributions differ. Thus, we select the mean and the standard deviation of the distribution as the features to distinguish attraction behaviors.

\section*{Clustering}
To discover common patterns of inflow within cities, regions are clustered using the attraction features discussed in the previous section. We used a Hierarchical Agglomerative Clustering (HAC) approach to categorize all $1492$ TAZes in Riyadh based on their attraction features. HAC classifies objects, where each object is represented as a vector of features that describe that object, based on specified similarity metrics. Here, a vector $x_i$ represent the attraction features that describe TAZ $i$ as follows:
\begin{equation}
x_i=[ inflow_i, SD_i, \mu_i, \sigma_i]
\end{equation}
Where $inflow_i$ is the inflow magnitude of TAZ $i$, $SD_i$  is the spatial dispersion of the inflow sources for TAZ $i$,  $\mu_i$ is the mean of the distances traveled to TAZ $i$, and $\sigma_i$ is the standard divination of the traveled distance distribution.

HAC starts by assigning each single object to a separate cluster, and sequentially merge the most similar clusters until it results in one cluster. Thus, HAC requires defining how to merge clusters and how to measure the distance between them. For merging clusters, We used complete-linkage algorithm, which merges two clusters based on their most dissimilar objects a follows:

\begin{equation}
D(X,Y)=\max_{x\in{X},y\in{Y}} d(x,y)
\end{equation}
Where $d(x,y)$ is the distance between two objects $x\in{X}$ and $y\in{Y}$, and $X$ and $Y$ are the 2 sets of clusters. Complete-linkage algorithm is conservative when merging clusters, thus it tends to find very compact clusters, which fits our objective in in finding closely related attraction patterns. For measuring the distance between clusters' objects $d(x,y)$, we use correlation based distance metric defined as follows:

\begin{equation}
d(x,y)=1-\frac{(x-\overline{x}).(y-\overline{y})}{{\lVert{(x-\overline{x})}\lVert}_2 {\lVert{(y-\overline{y})}\lVert}_2}
\end{equation}
Where $\overline{x}$ and $\overline{y}$ are the mean of the elements of vector $x$ and $y$ correspondingly, and $(x-\overline{x}).(y-\overline{y})$ is the dot product of the vectors $(x-\overline{x})$ and $(y-\overline{y})$. Correlation distance works well for finding unbalanced clusters sizes as we expect to have small number of places behaving very uniquely as strong attractors and larger number of places that are not as attractive. Additionally, the correlation score can correct for any scaling within a feature, while the final score is still being tabulated. Thus, different features that use different scales can still be used.

\begin{figure*}[t]
	\centering
	\frame{\includegraphics[width=0.6\columnwidth]{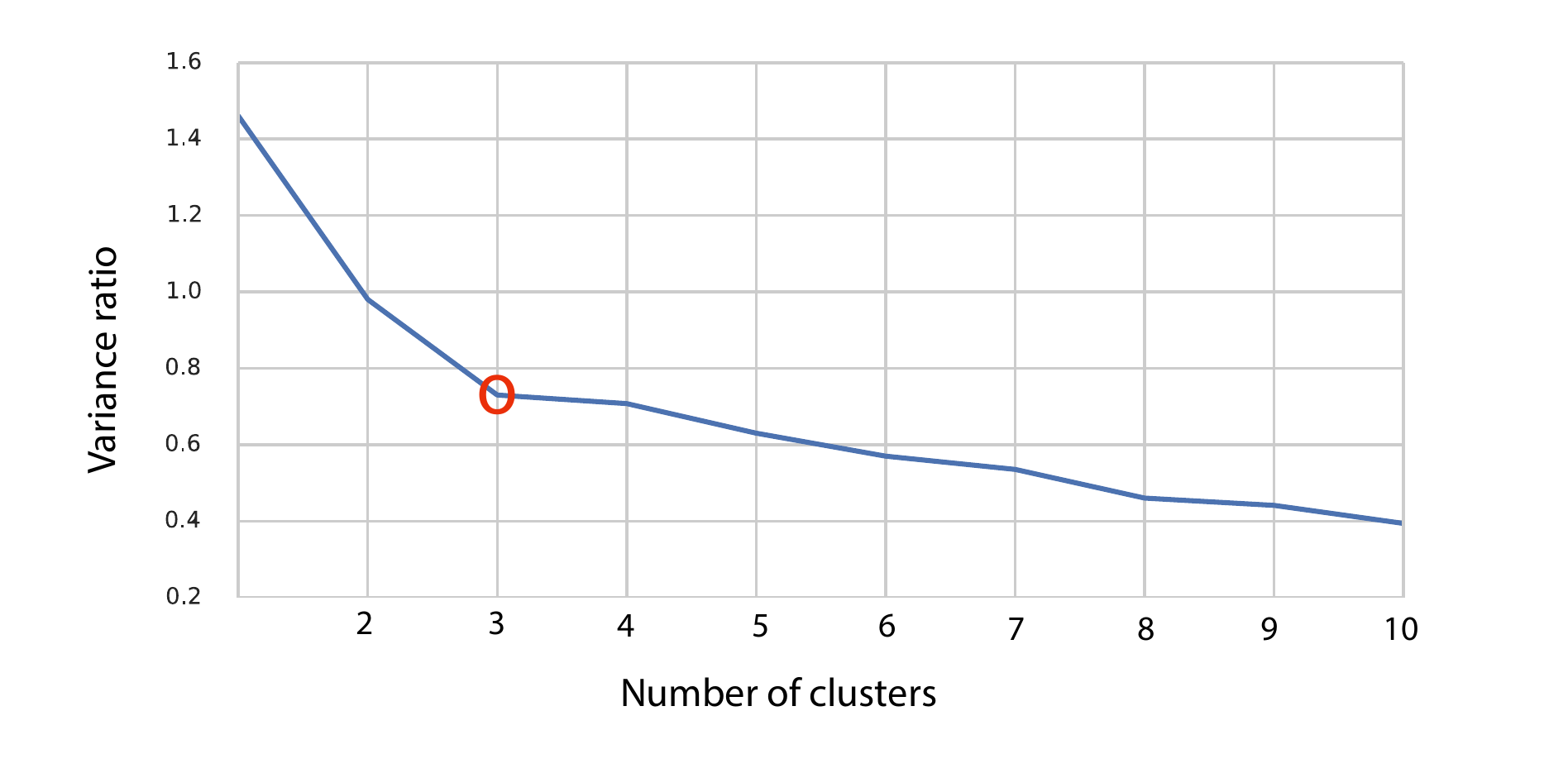}}
	\caption{The ratio of the within-cluster variance to the total variance for each possible choice of K (number of clusters). The variance decreases as the data is split into more clusters until it stops decreasing significantly at $k=3$ (marked by a red circle).}~\label{fig:figure7}
\end{figure*}

HAC provides a hierarchy structure of the classified regions in the city. To determine the number of clusters $k$ that best divide the data, we calculate the ratio of the between-cluster variance to the total variance for each possible $k$ from 1 to 10. The variance drops as $k$ increases, until it stops decreasing significantly. We select the $k$ that correspond to the point where the variance stops decreasing significantly, which is $k=3$ in our case as shown in Fig.{\ref{fig:figure7}}. The classification process over all TAZes in the city of Riaydh finds three types of attractors that have distinct features. The following section extends these findings to further interpret the results.

\begin{figure*}[t]
	\centering
	\frame{\includegraphics[width=1\columnwidth]{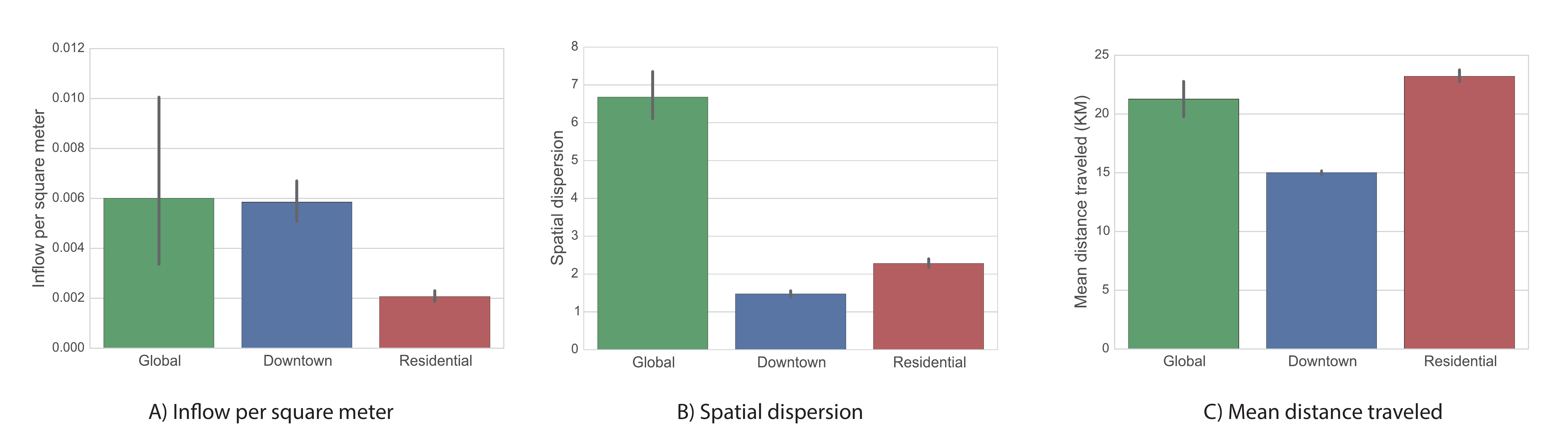}}
	\caption{The attraction features of the three detected types of attractors clusters. (A) shows the total inflow, which is high for both global and downtown attractors but extremely low for residential attractors. (B) shows the spatial dispersion of origins of visitors, where it's  significantly high in global attractors. (C) shows the mean of the distance traveled by visitors, where downtown attractors exhibit smaller distance mean due to its central location and thus higher accessibility to most visitors }~\label{fig:figure5}
\end{figure*}

\subsection*{Attractor Types}

Locations in Riyadh are classified into three main types of attractors based on distinguishable attraction of trips. Fig.\ref{fig:figure4} shows the 3 types of attractors classes detected, where the polygons are the TAZes. The global attractors are the ones that have significant influence on the whole city, hence their name. Unlike the remaining clusters, the locations of these places seem to be random around the city. The second detected type is the downtown attractors, which play a significant influence, after global, to attract trips. These are mostly clustered in the downtown area of the city. Finally, the residential attractors, are the least influential attractors in the morning period of typical weekdays. They are mostly located on the outer places of the city. 

\subsubsection*{Global Attractors} The most distinguishable feature of global attractors is the large spatial dispersion of the incoming flows, as visitors come from all over the city to visit these places, as shown in Fig.\ref{fig:figure5} A. Additionally, the amount of visitors they attract is the largest as shown in Fig.\ref{fig:figure6} B, where we use inflow per square meters due to the unbalanced sizes of the TAZes. Moreover, the mean distances traveled by visitors to these locations is extremely high, which makes these places highly attractive and unique. We call these places global attractors because they strongly influence human mobility over the whole city. Global attractors always offer some unique \textit{services} that makes them distinguished from other regions where visitors only find such services in those regions. Significant places in the city like the airport, major universities, and hospitals, that occupy TAZes on their own and are easy to identify from the map. Fig.\ref{fig:figure4} A shows annotation of these major places in the city.

\subsubsection*{Downtown Attractors} The second type of attractors is the downtown attractors, shown Fig.\ref{fig:figure5} B. They contain places that are mostly clustered in the central business district. These are TAZes that have relatively high inflow. However, because of their central location in the city, visitors from all over the city have short routes to access these places. They have smaller average distances compared to the other two types as shown in Fig.\ref{fig:figure4} C. For the same reason, the dispersion of the origins from the center of mass of inflows is also small as shown in Fig.\ref{fig:figure4} C. The significant feature of these places is that they attract great number of visitors and are accessible.

\subsubsection*{Residential Attractors} Residential attractors shown in Fig.\ref{fig:figure4} C attract the smallest number of visitors. As they are located in the outer sides of the city, the visitors to these places travel long distance on average to reach them as shown in Fig.\ref{fig:figure5} C. For the same reason, the dispersion of the small number of visitors is larger than the downtown attractors as shown in Fig.\ref{fig:figure5} B. 

\begin{figure*}[t]
	\centering
	\frame{\includegraphics[width=1\columnwidth]{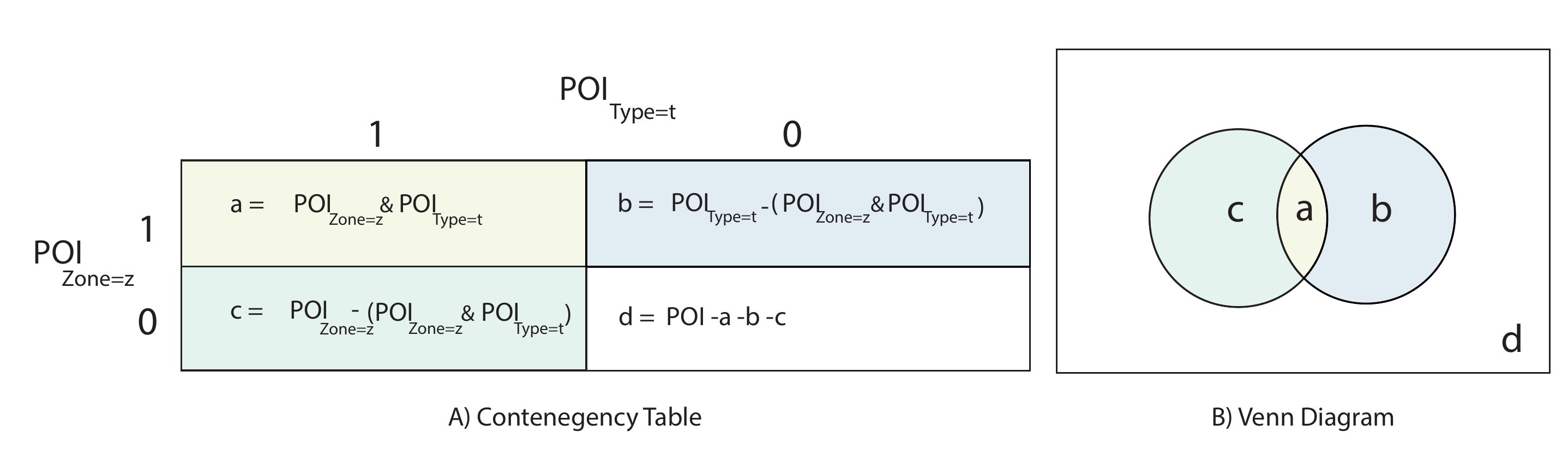}}
	\caption{The contingency table and the corresponding Venn diagram for statistical significance testing method.}~\label{fig:figure8}
\end{figure*}
\section*{Attractors and POIs}
We aim to further uncover what may cause the trips to the different types of attractors discovered. Thus, we relate the classified  TAZes to the composition of services offered in these TAZes. We used official Points of Interests (POIs) data that contains all places in Riyadh city, around $12,000$, offered by Arriyadh Development Authority (ADA), which is the official entity managing all urban planning tasks in the city. POIs are classified into 23 subtypes of services offered in the city such as restaurants, schools, hospitals ... etc. In the following section, We explain the methodology by which we quantify the relationship between different types of POIs and attractors types .  

\subsection*{Statistical significance testing}
We aim to relate each type of POIs to the different attractors types to explore what makes different places have different attraction profiles. Using only counts of POIs per attractor type is not enough, because the distribution of types of POIs is unbalance. For example, restaurants are very frequent and spatially distributed, whereas universities are much fewer and are present in few districts.  Thus, to identify which type of POIs are significantly concentrated in each type of attractors, we use a statistical significance testing approach. The statistical significance testing method measures the probability of observing the amount of a POI type in a spatial zone. It factors in the amount of all POIs in the spatial zone in addition to the amount of POIs with that type tag in the whole city. 

We used Fisher's Exact Test (FET) to relate each type of POIs to the different attractor types. We selected FET because it works for small observations and calculates the exact probabilities rather than approximations such as in Chi square test. FET aims to test the dependency between two categorical variables given the observed data. It takes a contingency table as an input, which represent the relationship between two categorical variables in terms of their frequency distribution and their overlap. In our context, the first variable $POI_{Type=t}$ is the number of POIs that belong to a specific type $t$ , and the second variable $POI_{Zone=z}$ is the number of POIs in a spatial zone $z$ . We aim to test the significance of the overlap $POI_{Zone=z} \& POI_{Type=t}$ ,which is the amount of POI $t$ in zone $z$, between these two variables. Fig.\ref{fig:figure8} shows the contingency table and the corresponding Venn diagram representation. In Fig. \ref{fig:figure8}, $a$ represents the overlap, which is the amount of POIs of type $t$ that are located in zone $z$, $b$ represents the amount of POIs of type $t$ that are not located in zone $z$, $c$ is the amount of other POI types located in zone $z$, and $d$ represents the number of all the rest of POIs in the city.

The objective is to test weather the amount of POI type $t$ is significantly concentrated in a spatial zone $z$. FET quantifies the probability (p-value) of observing that amount of POI type $t$ or larger in zone $z$ by chance. The smallest the p-value is, the strongest the concentration of type $t$ in zone $z$ is. FET calculates the p-value $p$ as follows:
\begin{equation}
p=\frac{(a+b)! (c+d)! (a+c)! (b+d)!}{a! b! c! d! n!}
\end{equation}
Where $n$ is the total amount of all POIs in the city. FET incorporates two crucial factors when measuring the significance. First, the significance depends on how many other POI types in zone $z$ as a measure of purity. If a zone has a large amount of POIs of type $t$ ,but it also has a lot of other POIs types, that makes POI of type $t$ less significance due to this impurity in the decomposition of all POIs in that zone. Second, the significance depends on the amount of POI of type $t$ that are not in the tested zone $z$ as a measure of rarity. If there are large number of POIs of type $t$ elsewhere, that makes type $t$ insignificance in the tested zone. These two features makes FET superior to trivial methods like calculating the percentages of POIs types in spatial zones.


\begin{figure*}[t]
	\centering
	\frame{\includegraphics[width=1\columnwidth]{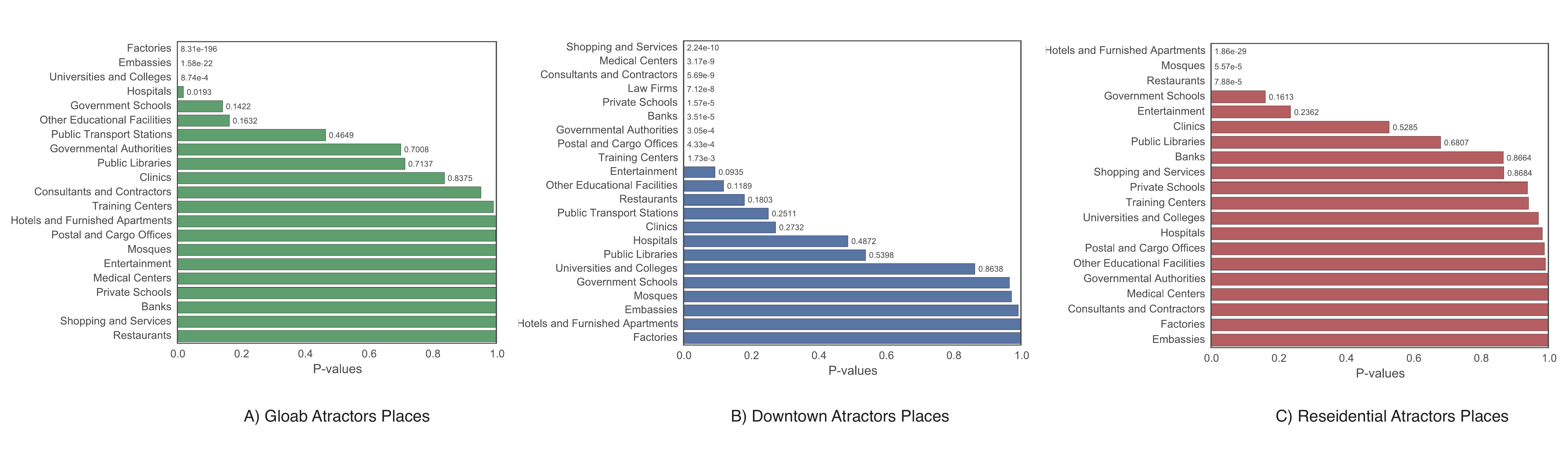}}
	\caption{The types of POIs ordered by their statistical significance in the corresponding attractor's type. (A) The p-values of finding each of the POI types in TAZes classified as global attractors, where POIs types are ranked by their significance showing that factories, embassies, universities, and hospitals are type of services significantly located in global attractors. (B) POI types significantly located in the downtown attractors, which mostly includes business types of services. (C) POI types significantly present in the residential attractors that includes services and amenities such as furnished apartments, mosques (places of worship for Muslims), and public schools. }~\label{fig:figure6}
\end{figure*}


We find that the proposed method can capture how each type of attractor has a different composition of POI types as shown in Fig. \ref{fig:figure6}. We discuss the attractors types and the significant POIs related to them in the following sections.
\subsubsection*{Global Attractors} Fig. \ref{fig:figure6} A shows each type of POIs ordered by their p-values, which represent how much they are concentrated in global attractors. Factories, embassies, universities, and hospitals are the top POI types that attract massive amounts of people coming from all over the city. The city has three major universities attracting a student body of $40k$ each contributing significantly to the observed global attraction. The city of Riyadh also has a major industrial area in the south to where a major number of factory workers commute. In addition, Riyadh city has the diplomatic quarter district, which hosts over $50$ embassies attracting workers and visitors from all over the city to process documents. Hospitals are also expected to cause global attraction as visitors come from all over the city.
\subsubsection*{Downtown Attractors}
Fig.\ref{fig:figure6} B shows the types of POIs mostly located in the downtown attractors in a descendant order by significance. The first observation is that we witness a large number of POI types with strong significance (p-values<0.01) compared to the other two types of at tractors. That is due to the richness in quantity and variation of types of POIs in this area. Thus, we expect larger number of POIs types to be significant in that area. The common theme of significant POI types in this attractor is businesses, typical in downtown regions.
\subsubsection*{Residential Attractors}
Fig.\ref{fig:figure6} C shows the types of POIs concentrated in the residential attractors in a descendant order. Most significant types of POIs are services needed in residential areas like furnished apartments, mosques (worship places for Muslims), restaurants, including small restaurants and fast food places, and public schools. These types of places are not unique, so each residential neighborhood has its own share of these places to serve the population living nearby. 

\section*{Conclusions and future directions}
We present a novel computational framework to discover different attraction patterns in cities. We proposed 3 dimensions to define attraction of urban zones: total number of incoming trips, the spatial dispersion of the origins of trips, and the distribution of distances traveled by visitors to reach that district. Further, we present a method for understanding the relation between the decomposition of the types of POIs in a spatial zone and its attraction behavior. We applied the method and discuss the results in the city of Riyadh, the capital of Saudi Arabia.

The results of implementing the discussed modules mine data from mobile phones to provide a coherent understanding of the dynamics of the interaction between the flows of people to a district and types of services (POIs) that are located in that district. We detect three attraction patterns in the city of Riyadh according to the morning mobility dynamics. Global attractors, receive large share of the visitors traveling longer distances and coming from all over the city. These attractors have places of interest that are the destination of large student bodies, factory workers, hospital associates, and embassies. The second type of attractor is that of the downtown area, which receives high inflow of people from smaller distances and spatial dispersion due to its central location in the city that makes them accessible. The most significant POIs types located in the downtown attractors are business based places like firms, shopping and service places. The least significant trip attraction is to the residential areas in the morning hours, where the amount of inflow is the lowest. Residential attractors contain common POIs that serve inhabitants such as apartments, mosques, and schools.
\\
Several interesting directions can follow this work. One is developing a predictive model of inflows and trip distributions, taking into account set of POIs, distance between origin and destinations and their population densities. Another possible direction is to compare these patterns among various cities to learn about best urban plans that reduce traffic.

\section*{Acknowledgments}
The research was supported in part by grants from the Center for Complex Engineering Systems at KACST and MIT,
and the U.S. Department of Transportation’s University Transportation Centers Program.

\nolinenumbers

%
%
%

\bibliographystyle{plos2015.bst}
\bibliography{UA}

\end{document}